\documentclass[reprint,amsmath,amssymb,aip]{revtex4-1}
\usepackage{lipsum}
\usepackage{graphicx}% Include figure files
\usepackage{dcolumn}% Align table columns on decimal point
\usepackage{bm}% bold math
\usepackage[utf8]{inputenc}
\usepackage[T1]{fontenc}
\usepackage{mathptmx}
\usepackage{etoolbox}

\usepackage[colorlinks=true, urlcolor=blue, linkcolor=blue, citecolor=blue]{hyperref}

\makeatletter
\def\@email#1#2{%
 \endgroup
 \patchcmd{\titleblock@produce}
  {\frontmatter@RRAPformat}
  {\frontmatter@RRAPformat{\produce@RRAP{*#1\href{mailto:#2}{#2}}}\frontmatter@RRAPformat}
  {}{}
}%
\makeatother
\begin{document}

\title%[Sample title]
{Optical realization of magneto-intersubband oscillations}
% Force line breaks with \\

\author{M.\,L.\,Savchenko}
\affiliation{Institute of Solid State Physics, Vienna University of Technology, 1040 Vienna, Austria}
%\affiliation{Rzhanov Institute of Semiconductor Physics, 630090 Novosibirsk, Russia}
\email{maxim.savchenko@tuwien.ac.at}

\author{A.\,A.\,Bykov}
\affiliation{Rzhanov Institute of Semiconductor Physics, 630090 Novosibirsk, Russia}
\affiliation{Novosibirsk State University, 630090 Novosibirsk, Russia}

\author{A.\,Shuvaev}
\affiliation{Institute of Solid State Physics, Vienna University of Technology, 1040 Vienna, Austria}

\author{A.\,K.\,Bakarov}
\affiliation{Rzhanov Institute of Semiconductor Physics, 630090 Novosibirsk, Russia}
\affiliation{Novosibirsk State University, 630090 Novosibirsk, Russia}

\author{A.\,Pimenov}
\affiliation{Institute of Solid State Physics, Vienna University of Technology, 1040 Vienna, Austria}

\author{O.\,E.\,Raichev}
\affiliation{Institute of Semiconductor Physics, National Academy of Sciences of Ukraine, 
03028 Kyiv, Ukraine}
%Prospekt Nauki 45

\date{\today}% It is always \today, today,
             %  but any date may be explicitly specified

\begin{abstract}%250 words
We report on the optical realization  of the magneto-intersubband oscillations that have been measured in the sub-terahertz transmittance of a GaAs quantum well with two subbands occupied. 
Following their dc analogue, the oscillations are periodic in the inverse magnetic field with the period governed by the subband gap.
%and do not have the temperature dumping.
%The oscillations follow presented dynamic simplified version of their dc analog, including the polarization dependence on the sign of the circular polarization. 
Their magnitude and polarization dependence 
%of the oscillations 
accurately follow the presented simplified
%ready-to-use 
version of the dynamic magneto-intersubband oscillations equation that naturally
combines dc magneto-intersabband oscillations with  microwave-induced resistance oscillations (MIRO).
Simultaneously measured photoresistance 
%$\delta R$ 
also reveals its strong sensitivity to the sign of the circular polarization, 
%when the maximum photoresistance ratio at positive and negative magnetic fields 
%$[\delta R(B)/\delta R(-B)]_{max} \approx 5$ 
%reaches 5, 
proving the used theoretical modeling.
%s of such family of photoinduced this and microwave-induced resistance oscillations (MIRO) phenomena. 
%with equal to 5 ratio at positive and negative magnetic fields equ
%of active and inactive.
\end{abstract}

\maketitle

%\section*{Introduction}\label{sec:intr}

Transport and optical measurements are complementary probes to study solid state structures. 
Despite being quite different in realization and in observed effects, they can be expected to result in one-to-one correspondence regarding the physical properties of studied systems. 
In contrast to a variety of transport experiments in heterostructures, several their dynamic counterparts still have to be detected and studied. 

\begin{figure*}
        \includegraphics[width=2\columnwidth]{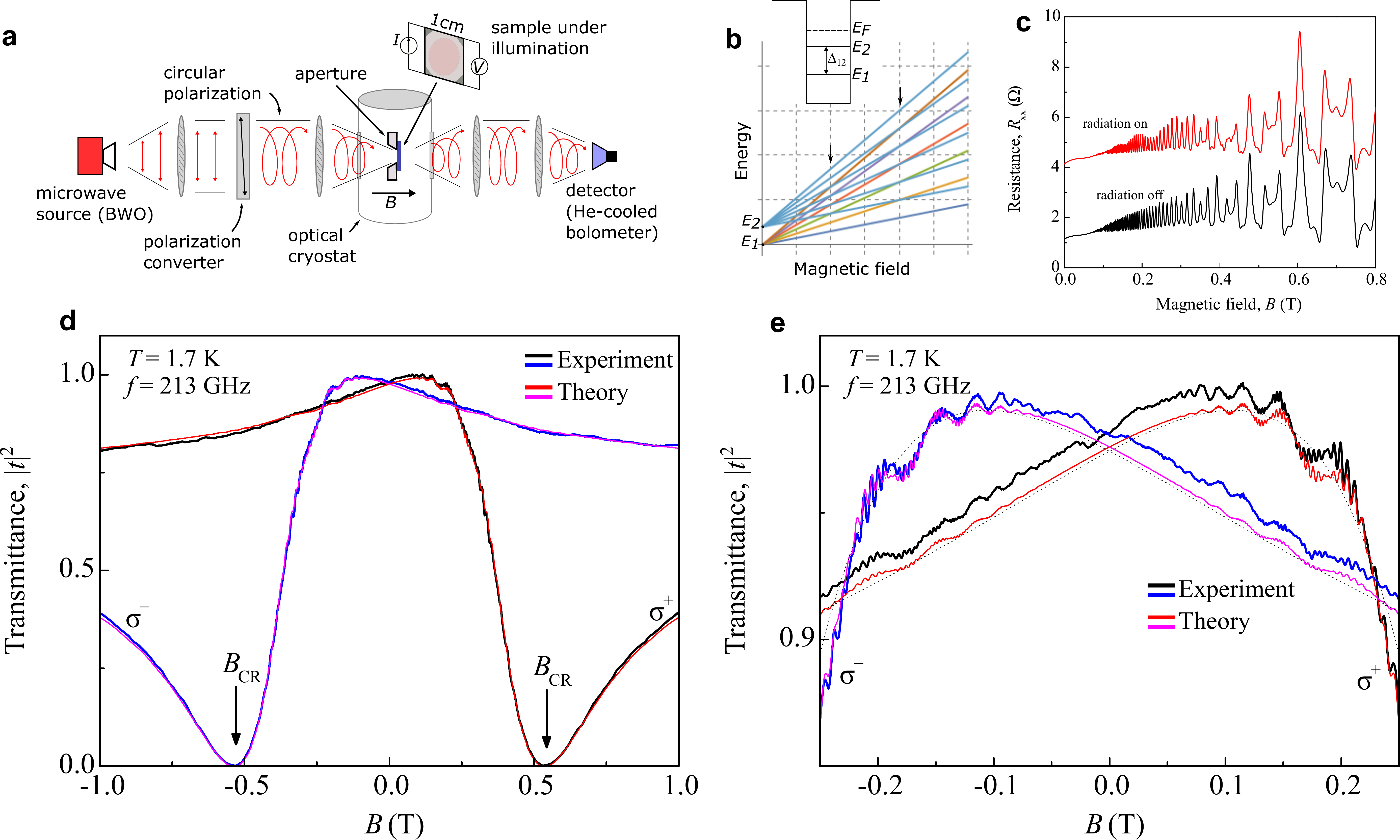}
        \caption{
        (a)~Experimental setup, optical and phototransport measurements were done in parallel. 
        (b)~A Landau level fan chart for a two-subband system, $E_1$ and $E_2$ represent bottoms of two subbands, the subband spacing $\Delta_{12} = E_2 - E_1$. 
        Arrows indicate the conditions of the resonant scattering between subbands when their Landau levels intersect.
        (c)~Magnetic field dependences of the measured resistance $R(B)$ when radiation is off (black) and on (red), the red curved is vertically shifted by 1.5\,$\Omega$ for clarity.
        Microwaves result in the additional modulation of the resistance oscillations.
        (d)~Magnetic field dependences of the transmittance $|t_\pm|^2(B)$ measured at $f = 213\,$GHz using right-hand (black, $\sigma^+$) and left-hand (blue, $\sigma^-$) circularly polarized radiation. 
        (e)~A zoom-in of panel (d) near zero field, where the studied oscillations are clearly seen.
        The theory curves are based on~Eqs.~\ref{eq: T}-\ref{eq: sD}, and \ref{eq: dSgima}.
        }\label{fig1}
\end{figure*}

One of such examples is given by the magneto-intersubband oscillations~(MISO), Fig.~\ref{fig1}\,(c). 
They are periodic in inverse magnetic field $B$ oscillations of the resistance $R$ 
of a high mobility quantum well with two populated subbands.
The frequency of these oscillations is equal to the subband separation over the cyclotron energy ratio, $\Delta_{12}/\hbar \omega_c$.
When this ratio is equal to an integer number, the Landau level sets of each groups intersect that results in the strong  scattering between the subbands, Fig.~\ref{fig1}\,(b).
%Each group of carriers form its own set of the Landau levels, $\hbar \omega_c (n+1/2)$ and $\hbar \omega_c (m+1/2) + \Delta$, and there is strong  scattering between subbands at the magnetic fields when the levels from different sets are crossing~(Fig.~\ref{fig1}\,(b)).
The prominent feature of MISO consists in the absence of their temperature damping. 
These resistance oscillations have been intensively studied both in theory~\cite{Polyanovskii1988, Raikh1994, Averkiev2001} and experimentally~\cite{Coleridge1990, Leadley1992, Sander1998, Rowe2001, Mamani2009}.

The presence of the external electromagnetic field modifies the magneto-intersubband oscillations, mixing them with the microwave-induced resistance oscillations~(MIRO)~\cite{Bykov2008a, Wiedmann2008, Raichev2008, Wiedmann2010a, Dmitriev2012}.
In its turn, the microwave-induced resistance oscillations represent the results of the optical transitions between distant Landau levels.
They inevitably modify the absorption 
%of the high mobility systems 
and have been thoroughly studied in optics~\cite{Abstreiter1976,Fedorych2010, Savchenko2020b}.
%Moreover, even in transport, it is becoming clear that the theory of microwave-induced oscillations can not explain all experimental findings~\cite{Fu2018, Dmitriev2019}.
However, the optical equivalent of MISO, when magneto-intersubband oscillations interfere with MIRO, was only presented in Ref.~\onlinecite{Bykov2015}, where due to technical limitations just a qualitative comparison with the theory~\cite{Raichev2008} was made.
%However, the combined effect of MISO and MIRO interference was only studied in dc transport with its optical realization. 
The optical response, as a more direct and 
simple method to analyze the radiation absorption, has helped to solve the long-standing 
problem of the polarization immunity of MIRO~\cite{Smet2005, Herrmann2016, Savchenko2022}.
However, at some conditions~\cite{Monch2022, Monch2023a}, the polarization immunity has been observed in a more general phenomenon, the cyclotron resonance absorption. This particular finding contradicts the basics of the light-matter interaction and still does not have even a qualitative explanation.
In this way, the optical realization of transport effects serves as a test for existing theories in the dynamic regime, and also establishes the static-dynamic correspondence in two-dimensional quantum systems.
This is especially important in connection with the growing interest and development of multilayer quantum materials and their applications in optical domain~\cite{Montblanch2023}.

%Qualitative agreement between the theory of Raichev and an experiment was obtained in~\cite{Bykov2015}.\\

Here we present an optical analog of the magneto-intersubband oscillations measured in direct transmission signal in a high-quality GaAs quantum well with two subbands occupied. 
The oscillations have been detected simultaneously in optical and dc transport responses. 
The optical oscillations and their polarization dependence provide an experimental proof for 
the theoretical predictions given in Ref.~\onlinecite{Raichev2008}.
%\textcolor{red}{And what to do with it?}

 %
%

%\section*{Methods}\label{sec:methods}

The studied GaAs quantum wells have been grown by molecular beam epitaxy on a GaAs substrate~\cite{Baba1983, Friedland1996, Umansky2009, Manfra2014}.  
The sample size was about 10$\times$10\,mm$^2$, the Ohmic contacts at the corners were used in the van der Pauw geometry and they were fabricated by burning indium droplets.
After exposure to the room light the total electron density and average mobility  were $n = 7.7\times10^{11}\,$cm$^{-2}$ and $\mu = 1.9\times10^6$\,cm$^2$/Vs, respectively. 
The optical measurements were performed in the Faraday geometry, see Fig.~\ref{fig1}, the sample was irradiated from the substrate side through a 8-mm metal aperture. 
A backward-wave oscillator was used as a source of the normally incident continuous monochromatic radiation. 
The transmittance through the sample was measured using a He-cooled bolometer. 
In parallel with the transmittance, the photoresistance $\delta R$ (the difference of the resistance signals in the presence and absence of irradiation) was measured using the double-modulation technique~\cite{Savchenko2020, Savchenko2020b, Savchenko2022}. 
All presented results were obtained at radiation frequency $f = \omega/2\pi  = 213\,$GHz and temperature $1.8$\,K.

In Fig.~\ref{fig1}~(c) we show the magnetic field dependences of the sample resistance, measured when the radiation source is switched off (black) and on (red) at $213\,$GHz right-hand ($\sigma^+$) circular polarization~(note the shift of the red curve).
Let us first discuss the magnetoresistance without illumination with sub-terahertz light. 
Several types of oscillations can be seen in these data.
First, high-frequency magneto-intersubband oscillations that survive up to the lowest magnetic fields. 
At higher magnetic fields they are supplemented by the Shubnikov -- de Haas oscillations from both subbands.
%\textcolor{red}{[Should we add anything known about these oscillations here?]}
These types of oscillations can be observed without sample irradiation. 
\begin{figure}
        \includegraphics[width=1\columnwidth]{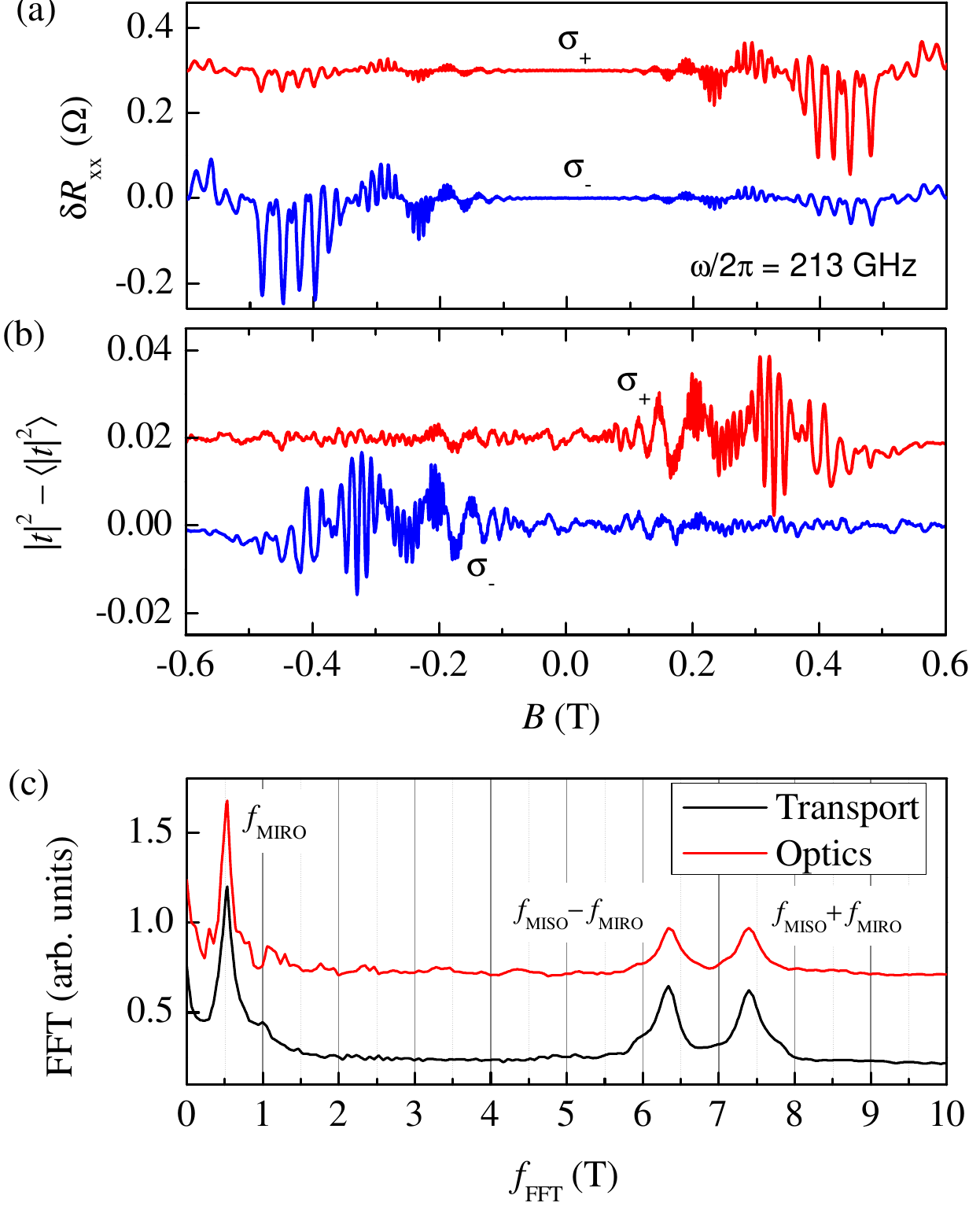}
        \caption{
        (a) and (b) Magnetic field dependences of photoresistance $\delta R(B)$ and transmittance oscillations $|t|^2 - \langle |t|^2 \rangle$ for right-hand (red, $\sigma^+$) and left-hand (blue, $\sigma^-$) circularly polarized radiation.
        A similar asymmetry for the curves at $\pm B$ is seen. 
        (c)~A fast Fourier transform (FFT) of the data from panel (a) and (b) for the right-hand circularly polarized radiation.
        The curves on all panels are vertically shifted for clarity.
        }\label{fig2}
\end{figure}
Low-power sub-terahertz radiation adds a kind of $\cos(2\pi\omega/\omega_c)$ modulation to the $R(B)$ dependence (red curve in Fig.~\ref{fig1}~(c)).  
This term 
%can come from different quantum~\cite{Dmitriev2012} and classical 
%may have different nature~\cite{Dmitriev2012}
%In this regime, 
originates from the radiation-induced transitions between distant Landau levels, which is the mechanism  responsible for the microwave-induced resistance oscillations in one-subband systems~\cite{Zudov2001, Dmitriev2012}.
Since the intersubband energy difference $\Delta \sim 10 $\,meV (see later for an exact number) is higher than the radiation energy $\hbar \omega \approx 0.9$\,meV, the MIRO-like modulation has a smaller frequency compared to the intersubband oscillations. 
%Moreover, the strength of this modulation is proportional to the radiation absorption rate that has maximum at the cyclotron resonance.
%Thereby, for the $\sigma^-$ polarization, when the cyclotron resonance is at negative magnetic fields, the radiation has nearly no affect on the $R(B)$ dependence.
%All these types of oscillations were already studied~[whome to cite?]
%\textcolor{red}{[Should we add anything known about these oscillations here?]}
 
%They are modulated by slower (since the intersubbands energy difference $\Delta$ is higher than the radiation energy $\hbar \omega$) sin(w/wc) factor related to the microwave-induced resistance oscillations (MIRO)~[Bykov's papers?] and originated from the radiation-induced transitions between distant Landau levels~\cite{Dmitriev2012}.
%The oscillation amplitude is higher at positive (negative) magnetic fields for the $\sigma^+$ ($\sigma^-$) polarization, following the strong radiation absorption in the cyclotron resonance conditions. 

 %Here the radiation was switched on \textcolor{red}{(we may remeasure the curve without rf)}.
 %The expected magneto-intersubband oscillations are seen by high-frequency oscillations on both curves.

In Fig.~\ref{fig1}~(d) we show the magnetic field dependence of the transmittance $|t|^2$ measured 
%at $f = 213\,$GHz and 
using the right-hand ($\sigma^+$) and left-hand ($\sigma^-$) circular polarizations.
Nearly zero transmittance at $|B_\text{CR}| \approx 0.54$\,T corresponding to the cyclotron resonance allows us to determine the cyclotron effective mass $m_{CR} \approx 0.071\,m_0$.
In Fig.~\ref{fig1}~(e) we show the expanded view near zero magnetic field in Fig.~\ref{fig1}~(d).
Two types of oscillations can be seen in these data as well. 
Their amplitude is higher at positive (negative) magnetic fields for the $\sigma^+$ ($\sigma^-$) polarization, having the maximum radiation absorption at the cyclotron resonance field. 
The lower-frequency oscillations are governed by the ratio $\omega/\omega_c$ and they essentially represent the dynamic analogy of MIRO that was studied in Ref.~\onlinecite{Savchenko2020b}. 
The higher-frequency oscillations are the optical realization of the magnetic-intersubband oscillations
that are the main result of our work.
%discussed in the Introduction Section. 

%Before analysing the transmittance oscillations, let's discuss photoresistance. 
Let us have a closer look at the observed oscillations in transport and optics.
The photoresistance $\delta R$ and the oscillating parts of the transmittance 
$|t|^2 - \langle |t|^2 \rangle$  are shown in Fig.~\ref{fig2}~(a) and (b), respectively.
Both curves are sensitive to the degree of the circular polarization (see the discussion below), and the corresponding FFT curves are presented in Fig.~\ref{fig2}~(c). 
From these data, it is evident that phototransport and transmittance reveal the same sets of oscillations, namely,  
the magneto-intersubband oscillations.
Thereby, we can use the knowledge on the photoresistance oscillations to analyse the FFT frequencies.

The lowest oscillation frequency comes from the microwave oscillations and is equal to $f_\text{MIRO} = B \omega/\omega_c = m \omega/e$, giving the quasiparticle effective mass value $m = 0.068m_0$.
This effective mass is about 4$\%$ lower than the cyclotron effective mass $m_{CR} \approx 0.071\,m_0$ that we attribute to the electron-electron interaction in the studied structure~\cite{Hopkins1987, Hatke2013, Kukushkin2015, Shchepetilnikov2017, Tabrea2020, Savchenko2020b}.
Two other FFT peaks are given by the combined frequencies, $f_\text{MISO} \pm f_\text{MIRO}$,
%, where $f_\text{MISO} = m\Delta /\hbar e$~\cite{Bykov2017}.
%Obtained from $f_\text{MIRO}$ electron effective mass is equal to $m = 0.068 m_0$ and subband energy splitting is equal to $\Delta = $.
since just transport intersubband oscillations have nearly no sensitivity to temperature variations~\cite{Raichev2008} and, hence, are not seen in photoresistance.
However, since the scattering rate that governs the system conductivity has an 
intersubband contribution, the MIRO-induced photoresistance has this type 
of contribution as well.
The formation of the combined frequencies $f_\text{MISO} \pm f_\text{MIRO}$ represents 
the interference of MIRO and MISO effects.
%modified by intersubband transitions scattering rate
%intersubband oscillations are governed by the subband density difference, 
%These intersubband oscillations are governed by the subband density difference, 
%Taking into account the $\cos(2 \pi \Delta/\hbar \omega_c) = \cos(\pi h|n_1 - n_2|/e B)$ modulation of MISO, 
%$\cos\frac{2 \pi \Delta}{\hbar \omega_c} = \cos\frac{\pi h|n_1 - n_2|}{e B}$
%These intersubband oscillations are additionally modulated by the radiation-induced transitions between distant Landau levels, which, in the simplest case, has a $\cos(2 \pi \omega/\omega_c) = \cos(2 \pi \omega m/e B)$ dependence.
%The periodicity of the CR harmonics gives the quasiparticle mass $m = 0.068\,$m$_0$, which value is typical for electrons in GaAs quantum wells~[?].
%In this way, we obtain from the FFT analysis the effective mass 
The periodicity of the MISO governed by $\cos(2 \pi \Delta_{12}/\hbar \omega_c) = \cos(\pi h|n_1 - n_2|/e B)$ allows us to find the difference in subband densities $n_1 - n_2 = 3.3\times10^{11}\,$cm$^{-2}$.
Considering the total density from the Hall measurements equal to $7.7\times10^{11}\,$cm$^{-2}$, we get $n_1 = 5.5\times10^{11}\,$cm$^{-2}$ and $n_2 = 2.2\times10^{11}\,$cm$^{-2}$.
These values coincide with the SdH analysis of the transport data, proving our model. 

%\subsubsection*{Transmittance oscillations}
Now, let us analyze the transmittance oscillations quantitatively.
The magnetic field dependence of the transmittance 
%overall shape of the 
$|t(B)|^2$  
%for the transmittance 
of the circularly polarized light through a dielectric slab containing an isotropic 2DES can be fitted using a standard equation~\cite{Abstreiter1976, Savchenko2022}
\begin{equation}
\label{eq: T}
%|t|^2  = \frac{1}{|1 + \sigma Z_0/2|^2},
%|t_\pm|^2  = \frac{1}{|1 + \sigma_\pm Z_0/2|^2},
|t_\pm|^2  = \frac{1}{|s_1(1 + \sigma_\pm \,Z_0) + s_2|^2},
\end{equation}
%
%\textcolor{blue}{After Eq. (1):}\\
where $\sigma_{\pm}$ is the dynamic conductivity for the two-subband system 
(which is specified later),
plus and minus signs correspond to the right and left-handed circular polarization, respectively,
$Z_0 \approx 377\,\Omega$ is the  impedance of vacuum, and
%, and $e$ is the elementary charge.
two complex parameters $s_1 = 1/2(\cos(kd) - i \epsilon^{-1/2} \sin(kd))$, $s_2 = 1/2(\cos(kd) - i \sqrt{\epsilon} \sin(kd))$ describe the Fabry-P\'{e}rot interference in the substrate, and are controlled by 
%the product of 
the wave number $k$, the thickness $d$
and the dielectric constant $\epsilon$ of the substrate, respectively.
%$k=\sqrt{\epsilon}\,\omega / c$, $\epsilon$ is the dielectric constant of the substrate. 
%and the dielectric constant of the substrate $\epsilon$. 
%Note that in case several charge carriers coexist in 2DES, the dynamic conductivity in Eq.\,(\ref{eq: TD}) is given by the sum of respective contributions.
%\textcolor{blue}{The paragraph including the other equations:}\\

To reproduce the oscillations in the measured $|t|^2(B)$ dependence, we calculate $\sigma_{\pm}$ taking into account the quantum correction to the conductivity due 
to the Landau quantization. 
Following Ref.~\onlinecite{Raichev2008}, away from the cyclotron resonance, when $\mu |B_{CR} \mp B| \gg 1$, we obtain
%\footnote{I prefer to retain here the exponential form in $\delta \sigma$, as it emphasizes that $\delta \sigma$ has both real and imaginary parts, though the relative contribution of the imaginary part is small and further neglected. In this form $\delta \sigma$ follows from Ref. [1], while in the single-subband limit reproduces the result of I. A. Dmitriev, A. D. Mirlin, and D. G. Polyakov, PRL 91, 226802 (2003).}:
%
\begin{gather}
\sigma_{\pm} =\sigma^D_{\pm} +\delta \sigma_{\pm}, \label{eq: s+ds}\\
\sigma^D_{\pm}=i \frac{e n}{B_{CR} \mp B} + \frac{n m}{(B_{CR} \mp B)^2} \nu_0, \label{eq: sD} \\
\delta \sigma_{\pm}=\frac{2 n m}{(B_{CR} \mp B)^2} \delta \nu \exp \left( i \frac{2 \pi \omega}{\omega_c} \right),  \nonumber
%2
\end{gather}
%
%\begin{gather}
%\sigma_{\pm} =\sigma^D_{\pm} +\delta \sigma_{\pm}, \nonumber \\
%\sigma_{\pm}=i \frac{|e| n}{B_{CR} \pm B} + \frac{n m}{(B_{CR} \pm B)^2} \nu_0, \nonumber \\
%\sigma^D_{\pm} =\frac{n m}{(B_{CR} \mp B)^2} \nu_0 - i \frac{e n}{B_{CR} \mp B}, \nonumber \\
%\delta \sigma_{\pm} = 2 {\rm Re} \sigma^D_{\pm}\,\delta\!\nu \exp \left( i \frac{2 \pi \omega}{\omega_c} \right),  
%\nonumber
%\end{gather}
where $e = |e|$ is the elementary charge and
\begin{eqnarray}
\nu_0=\nu^{tr}_{11} \frac{n_1}{n} + \nu^{tr}_{22} \frac{n_2}{n} + \nu^{tr}_{12},
\end{eqnarray}
\begin{eqnarray}
\nonumber
\delta\!\nu=\delta^2_1 \nu^{tr}_{11} \frac{n_1}{n} + \delta^2_2 \nu^{tr}_{22} \frac{n_2}{n} +\delta_1 \delta_2 
\nu^{tr}_{12} \cos \frac{2 \pi \Delta_{12}}{\hbar \omega_c}.
\end{eqnarray}
In these expressions, 
%$\sigma^D_{\pm} = en\mu/(1 - i\mu(B_{CR} \mp B))$ is the Drude conductivity, $\mu$ is the static mobility
%\textcolor{red}{[Oleg, or should we rewrite it through $\nu_0?$]}, 
$\nu^{tr}_{jj'}$ are the transport scattering rates for the 
transitions between subbands $j$ and $j'$, $\omega_c = e|B|/m$, 
%$m$ is the quasiparticle effective mass, 
%cyclotron frequency that is determined by the 
$\delta_j = \exp(-\pi \nu_j/\omega_c)$ are the Dingle factors expressed through 
the quantum scattering rates $\nu_j=\nu_{jj}+\nu_{12}$.
%, and $\Delta_{12}=\pi\hbar^2 |n_1-n_2|/m$ is the subband separation energy. 

In order to compare the theory to experimental data, we simplify the equation by 
assuming that the transport and quantum relaxation rates are the same for both 
subbands: $\nu^{tr}_{11}=\nu^{tr}_{22}$ and $\nu_{11}=\nu_{22}$.
This assumption is justified since there is no strong classical magnetoresistance %in the classical region of $B$
(Fig.~\ref{fig1}\,(c)), otherwise the simplified theory would not work, and the magnetoresistance itself can be used to extract the parameters~\cite{Mamani2009}.
%The last equality means 
These assumptions also mean
that the Dingle factors are equal as well, $\delta_1=\delta_2 \equiv \delta = \exp(-\pi /\mu_q |B|)$, where $\mu_q$ is the quantum mobility. 
%\textcolor{red}{We also neglect the difference between intersubband quantum and transport rates:  $\nu_{12}=\nu^{tr}_{12}$. The last approximation is justified when the main mechanism of intersubband scattering comes from a short-range correlated impurity potential, which is reasonable in view of large difference of the Fermi momenta in the subbands [we don't need this assumption, instead, I used $\mu_q$ as an independent fitting parameter].}
In these approximations, the rate $\nu_0=\nu^{tr}_{11} + \nu^{tr}_{12}$ 
coincides with the transport relaxation rate $\nu_{tr}$ entering the static 
mobility, $\mu = e/m \nu_{tr}$; otherwice the connection between $\nu_0$ and 
$\nu_{tr}$ is more complicated~\cite{Mamani2009}. 
%[N. C. Mamani et. al., Phys. Rev. B 80, 085304 (2009)]
%\textcolor{red}{here I suggest to cite this paper}. 
Finally, neglecting the imaginary part of $\delta \sigma_{\pm}$ in view of its relative 
smallness, ${\rm Im} \delta \sigma_{\pm}/{\rm Im} \sigma_{\pm} \ll {\rm Re} 
\delta \sigma_{\pm}/{\rm Re} \sigma_{\pm}$, we obtain the final expression for the conductivity correction related to the magneto-intersubband oscillations
\begin{eqnarray}
\label{eq: dSgima}
\delta \sigma_{\pm} = 2{\rm Re} \sigma^D_{\pm}\, \delta^2 \cos \frac{2 \pi \omega}{\omega_c}
\left(1 + \gamma \cos \frac{2 \pi \Delta_{12}}{\hbar \omega_c}  \right),
\end{eqnarray}
where $\gamma=\nu^{tr}_{12}/\nu^{tr}_{jj}$ denotes the ratio of intresubband and 
intrasubband scattering rates, which is considered below as an adjustable parameter.
%\textcolor{red}{I do not like the notation $\nu^{tr}_{0}$, it mixes with $\nu^{tr}_{jj'}$ and, as you see, it is not necessary to introduce this notation.} 
In the case of only one subband being occupied (or there is no scattering between subbands) there are no intersubband oscillations, the second term in the brackates is equal to zero, and the remaining $\cos(2 \pi \omega/\omega_c)$ oscillations represent the quantum, MIRO-like, oscillations of transmittance that were studies in Ref.~\onlinecite{Savchenko2020b}.
If there are two subband occupied but there is no radiation applied, the first cosine is equal to unity and we 
obtain standard transport magneto-intersubband oscillations.
If both factors are relevant, one can detect the dynamic magneto-intersubband oscillations~(Fig.~\ref{fig2}).

To fit the experimental $|t|^2(B)$ data shown in Fig.~\ref{fig1}~(d) and (e) we combine the Drude conductivity $\sigma^\text{D}_\pm$, Eq.~\ref{eq: sD} with its correction $\delta\sigma_\pm$ from Eq.~\ref{eq: dSgima},
and insert the obtained total conductivity of the system to Eq.~\ref{eq: T}.
%to get the total conductivity of the system.
%By combining the Drude conductivity $\sigma^\text{D}_\pm$ and its correction $\delta\sigma_\pm$ from Eq.~\ref{eq: dsigma_final} one can get the total conductivity of the system.
%Inserting this to the transmittance Eq.~\ref{eq: TD} allows us to fit the experimental data
Two fitting parameters, the quantum mobility $\mu_\text{q} = 2 \times 10^5~$cm$^2/$Vs and the relative strength of the intersubband scattering $\gamma = 0.4$, allow us to obtain a very 
good fit of the transmittance, which additionally proves the validity of the simplifications.

%\subsubsection*{Polarization dependence of photoresistance}

Now we will turn to the observed polarization dependence of the measured photosignal.
In Fig.~\ref{fig2}~(a) we show the magnetic field dependences of the photoresistance measured at two circular polarizations.
This polarization dependence is governed by the polarization dependence of the cyclotron absorption and does not depend on the exact mechanism of the oscillations\cite{Dmitriev2012, Savchenko2022}.
Thus, the ratio of the photoresistance at positive and negative magnetic fields should be equal to the corresponding ratio of the cyclotron absorption, $\delta R(B)/\delta R(-B) = A(B)/A(-B)$.
The absorption can be calculated as $A = Z_0 |t_\pm|^2 \text{Re}\,\sigma_\pm^\text{D}$, therefore, 
the photoresistance ratio for the first oscillations harmonics should be equal in our case to about 18~(at $B \approx 0.43~$T).
%the maximum amplitude of the first oscillations harmonics should be equal in our case to about 18~(at $B \approx 0.43~$T).
%\textcolor{red}{[alternatively, we can estimate the quality of the polarization -- what is the percent of an opposite polarization that would result in a detected ratio?]}
The experimentally obtained ratio is equal to only about 4, which most probably comes 
from polarization distortions due to metallic aperture. As has been investigated elsewhere~\cite{Savchenko2022}, 
the boundary conditions at the aperture lead to a local admixture of a linear polarization, which finally makes 
the $\delta R(B)$ dependence more symmetric. 
The experimentally observed asymmetry can be
explained by 88$\%$ to 12$\%$ admixture of right and left circular polarizations, respectively.

%\section*{Conclusion}

In this work we demonstrated the optical analogy of the magneto-intersubband oscillations in the sub-terahertz transmittance signal.
The detected oscillations precisely follow the theoretical predictions extended to the dynamic domain.
This work, together with Refs.~\onlinecite{Savchenko2020b, Savchenko2024}, 
consistently demonstrates and proves the ability to detect and study the high-frequency dissipative transport features using contactless optical methods.
Moreover, the dynamic oscillations reveal another advantage of studying the optical response to simultaneously get the information about the Drude parameters and their corrections by electron-electron correlations.
Taking into account that the optical approach does not require high-quality structures and any specific fabrication processing, it can be used to test a variety of 2DES, including multi-valley systems and the steadily growing family of two-dimensional materials.

The observed strong polarization dependence of the photoresistance in a sub-THz frequency range together with the observation of the corresponding oscillations in the transmittance additionally prove the validity of the used theoretical models used in Ref.\,\onlinecite{Raichev2008}. 

%\subsubsection*{Outlook} 
%\textcolor{red}{[Maybe to say something about our predictions what else can be studied using such optical setup]}\\
While presented experiments only prove the theoretical predictions, there are established in dc transport effects that do not have high-frequency extensions yet.
First, the study of the transmission in the conditions of zero resistance state (ZRS)~\cite{Dmitriev2012}, when 4-point resistance is equal to zero. 
The current and not yet full understanding of this phenomenon~\cite{Dmitriev2019} is based on the domain structure, when the local conductivity is either positive or negative that can be spontaneously switched~\cite{Dorozhkin2011}. 
The switching leads to the breaking of the Drude approach, 
making such a study especially promising.
Another possible direction for investigations is related to the optical realization of the weak localization effects~\cite{Ihn2010}.
Here, the dc studies of the effect in the absence and presence of the radiation were performed~\cite{Vitkalov1988, Bykov1988}. 
However, the optical measurements as well as the dynamic extension of equations are still missing.
Finally, the recently discovered polarization immunity problem of the cyclotron absorption observed at some conditions~\cite{Monch2022} is also waiting for its explanation.
All these and other studies of subtle optical properties of 2DES require a solid understanding of experimentally proved theories  that,  among the other things, we present in this work.
%\textbf{We also note that we omit here the Shubnikov -- de Haas like term that is linear by $\delta$ but is suppressed at these temperatures.}

\section*{DATA AVAILABILITY}

The data that support the findings of this study are available from the corresponding author upon reasonable request.

\section*{REFERENCES}

% Create the reference section using BibTeX:
\bibliography{library}

\end{document}